\documentstyle[prl,aps,preprint,epsf]{revtex}
\begin{document}
\draft
\tightenlines
\preprint{GTP98-4}
\title{Transitions to quantum chaos in a generic one-parameter 
        family of billiards}
\author
{ Sunghwan Rim and Soo-Young Lee }
\address
{ Basic Science Research Institute, Korea University,
Seoul 136--701, Korea}
\author
{Eui-Soon Yim}
\address
{  Department of Physics, Semyung University, Chechon 390-230, Korea
}
\author
{C. H. Lee}
\address
{ D\&S Dept.,R\&D Center, Anam Semiconductor Inc., Seoul 133-120, Korea 
}
\date{\today}
\maketitle


\begin{abstract}
        Generic one-parameter billiards are studied both classically
        and quantally. The classical dynamics for the billiards
        makes a transition from
        regular to fully chaotic motion through intermediary soft chaotic
        system.
        The energy spectra of the billiards are computed 
        using finite element
        method which has not been applied to the euclidean billiard.
        True generic quantum chaotic transitional behavior and
        its sensitive dependence on classical dynamics are uncovered
        for the first time. That is, this sensitive dependence
        of quantum spectral measures
        on classical dynamics is a genuine manifestation of quantum chaos.
\end{abstract}

\vspace{0.5cm}

 
\vspace{1cm}


	Since the pioneering work of McDonald and Kaufman\cite{Mc79,Mc88},
        the study of 
	statistical measures of quantum spectra has been one of the major
        themes for quantum chaos.
        This study uncovered universal behaviors for integrable
        and  most of hard chaotic systems. 
        For transitional systems, i.e., from integrable to hard
        chaos through soft chaos senario,
        however, few observations have  been
        made\cite{Ro84,Ha84,Is85,Se84}.
        Among those observations, Robnik's\cite{Ro84,Ro93} analysis showed 
        using level spacing statistics that
        there is a gradual continuous transition from regular to hard chaos.
        H\"{o}nig 
        and Wintgen's analysis\cite{Ho89}, on the other hand,
        discovered that level spacing statistics
        showed extraordinary behavior in the case of
        the corresponding classical 
        dynamics with complexity, and suggested that the 
        understanding of it require
        a detailed knowledge of the underlying classical dynamics rather than
        a knowledge of the global classical phase space structure. 
        Grasping a generic transitional quantum behavior is very
        important to understand quantum chaos itself which has not
        been defined very clearly.
        These themes motivate this study.

        In this Letter we present the results of complete analysis, both
        classically and quantally,
        of a generic one-parameter family of
        billiard systems\cite{rim96,lyl98}.
        As the generic one-parameter billiards, we studied Dreitlein's
        billiards\cite{Dt89}
        whose boundary consists of two parallel lines with their 
        separation as length and  
        arcs of circles of a radius at both extremes.
        Fig. 1 shows a typical Dreitlein's billiard.
        If we vary radii of the circlular arcs the system changes from
        a square billiard to the Bunimovitch stadium.
        We parametrize the billiards by a parameter $\lambda$ as follows,
      \begin{equation}
         \lambda = \exp\, (-\sqrt{R^{2}-1} \,),
      \end{equation}
        where $R$ is the radius of circular arcs.
        As we change $\lambda$ from $0$ to $1$, Dreitlein's billiards
        change from a square billiard to the Bunimovitch stadium.
        At some parameter value
        $\lambda = \lambda_{c}$, the billiards become a hard
        chaotic system, i.e., K-system, through soft chaos.
        Non-analyticity of the boundary for Dreitlein's billiard
        makes its classical dynamics relatively
        simple compared to the other one-parameter family of billiards with
        analytic boundaries.
        This simplicity helps
        to get the complete analysis of its classical dynamics
        by studying the
        Birkhoff surface of sections, linear stability of periodic orbits,
        and imbedded island structures.

        We summarize the results of our analysis of classical dynamics
        of Dreitlein's billiards in the following.
        As the parameter $\lambda$ increases from $0$ to up, 
        the billiard starts to deform
        from a square billiard. Its Birkhoff surface of section
        (illustrated in Figure 2) shows lots of islands contiguous
        with each other.
        The islands of stability, surrounding elliptic fixed points, 
        are divided
        by a very thin connected chaotic region.
        The absence of separatrix motion due to non-analyticity of the boundary
        makes its 
        structure of the surface of section simple, i.e.,
        there is only one major connected chaotic region.
        As we increase $\lambda$ further, most of the islands quickly disappear.
        The most robust islands surround the 
        corresponding period two and period four orbits.
        Our present numerical analysis shows 
        that the island which surrounds period four orbit undergoes
        period doubling bifurcation at $\lambda \!=\! 0.167$. It is in good
        agreement with $\lambda \!=\! e^{-1-2^{-1/3}} \!=\! 0.1663...$
        obtained from the study
        of monodromy matrix.
        We numerically observed that small isolated regular regions from
        period doubling bifurcation disappear in the chaotic sea
        at about $\lambda = 0.198$
        The most robust island,
        which surrounds period two orbit,
        is found to disappear at
        $\lambda_{c} \!= 0.368$. The study of monodromy matrix produces
        $\lambda_{c} \!=\! e^{-1}$ which gives very excellent agreement.
        After that, Dreitlein's billiards become fully chaotic system
        (hard chaotic system) as $\lambda$ increases.

	Quantum mechanical analysis should solve the Schr\"{o}dinger equation
        with Dirichlet boundary conditions.
        This is equivalent to solve the Helmholtz equation, for which 
        finite element method(FEM) is employed. 
        FEM is, in principle, one of the best method to deal 
        with the boundary problems. The 
        advantages of FEM are known to be able to obtain both eigenvalues and
        eigenfuntions at a time, and to provide solutions to many complicated
        problems that would be intractable by other techniques. 
        Our FEM approach is the first application to the euclidean quantum
        billiard, to
        our best knowledge. 
        We normalize the area of the billiards
        to $\pi$ so that the mean level spacing is
        the same for all Dreitlein's billiards.
        Since the classical chaoticity comes from the circular arcs of
        Dreitlein's billiards,
        the boundary approximation is very important to obtain accurate
        results.
        Because of moderate curvatures of Dreitlein's billiards,
        iso-parametric coordinates give very good approximation.
        According to Henshell's estimate\cite{hs76}, 
        the typical radial errors are
	within less than $10^{-5} \%$.
        By using FEM, we calculate odd-odd parity
        eigenvalues and eigenfunctions of Dreitlein's billiards up to
        about $2400$th states
        for each of $46$ different parameter values of $\lambda$.
        Analysis of these data provides the ways to investigate 
        level spacing statistics,
        spectral rigidities and patterns of eigenfunctions.

        In this Letter, we are limited to discuss 
        only the level spacing statistics.
        The study of the other subjects will be followed near future.
        The first $1100$ reliable eigenvalues are used for each
        $\lambda$ to obtain 
        accurate level spacing statistics.
        To study level spacing statistics of Dreitlein's billiards,
        we tried to fit the data to the
        Brody distribution
        which is originally suggested by Brody $et. al.$\cite{br81}
        and later used by Robnik\cite{Ro84}.
        The Brody distribution interpolates Poisson and Wigner
        distributions which are characteristic
        level spacing distributions of integral
        and chaotic spectra, respectively.
       \begin{equation}
          P(s)  =  a \, s^{\nu} \exp \,(-b \,s^{\nu +1} \, )
       \end{equation}
          where $a$ and $b$ are determined by normalization conditions, i.e.,
        $\int_{0}^{\infty} P(s) \,ds \,=\, 1$ and 
        $\int_{0}^{\infty} s \, P(s) \, ds \,=\, 1$.
        The level spacing exponent $\nu$ is a measure of short range 
        interaction between energy levels, i.e., level repulsion.
        One of advantages to use the Brody distribution is that
        we can define cumulative distribution\cite{Mc88}.
        Since it is a smoother function of $s$ than $P(s)$, it is more
        easily fitted by data.
        We fitted our data to the cumulative Brody distribution,
        $W(s) \,=\, \int_{0}^{s} P(x) \, dx$,
        in order to obtain best fitted $\nu$.
	Fig. 3 shows our main results of the study.
        There are three different
        parameter regions which show
        different qualitative behaviors.
        We call these regions Region I, II and III.

        In the soft chaotic region (Region I),
        i.e., $0 <\! \lambda \!< \lambda_{c}$,
        it clearly shows that the distribution is moving from a Poisson to
        a Wigner distribution.
        However, it is certainly not a gradual continuous transition.
        As soon as $\lambda$ increases from $0$, $\nu$ increases sharply
        and reach about $0.6$ when $\lambda$ becomes $0.05$. And then it
        tends to saturate until $\lambda$ becomes $0.175$.
        Notice that the lowest value of $\lambda$ in that stretch
        is obtained surprisingly at $\lambda = 0.167$, where
        period two orbits lose their stability.
        There is a big jump in $\nu$ at $\lambda = 0.175$ where
        there is only one major isolated integrable region left in the classical
        phase space.
        After that $\nu$ fluctuates again
        around $0.7$ for a while and followed by
        another big jump at $\lambda = 0.36$ which is again very
        close to $\lambda_{c} = 0.368$ where the system becomes
        completely chaotic, i.e., K-system.
        The observation in Region I implies that quantum chaotic transition
        in Dreitlein's billiards
        is very sensitive to the corresponding classical dynamics.
        This reflects that there is an intrinsic coincidence between
        quantum chaos and classical dynamics.
	In the hard chaotic region (Regions II and III),
        i.e., $\lambda_{c} <\! \lambda \!< 1$,
        we can see two different behaviors depending on $\lambda$.
	For the region $\lambda_{c} <\! \lambda \! < 0.85$,
        as we expected, the level spacing distribution is close to the 
        Wigner distribution, which is believed to be 
        a universal characteristic behavior
        of hard chaotic system. In Region II, $\nu$ is bigger than $0.85$ except
        at $\lambda \!=\! 0.7$.
        This implies strong level repulsions between levels.
        However for the region $0.85 <\! \lambda \! < 1$, 
        the exponent $\nu$ shows a clear
        tendency to decrease.
        This is an unexpected results which needs an explanation.
        The explanation of this behavior will be the main topic of our 
        future work.
        Finally, we would like to mention that the exponent $\nu$ at 
        $\lambda \!=\! 1$ is $0.717$ which is in good agreement with
        McDonald and Kaufman's $0.71$ for the Bunimovitch stadium.
	Through our complete analysis of the corresponding classical system,
        we have found that quantum spectral measures are sensitively
        related to the topological changes of
        phase space manifolds of the classical system.

	In conclusion in the present Letter, 
        the major finding in this study is that the quantum chaotic transition 
        is not a smooth gradual transition.
	There are big changes in the distribution of eigenvalues for
        small changes in $\lambda$.
        It should be noticed that the parameter ranges,
        $0.05 <\! \lambda \!< 0.167$ and $0.175 <\! \lambda \!< 0.36$, 
        correspond to parameter ranges in which classical phase spaces 
        have two major isolated islands and one major
        isolated island, respectively.
        And the transition values, $\lambda \!=\! 0.175$ and
        $\lambda \!=\! 0.36$, are very close to
        the values at which period four and period two orbits lose
        their stability.
        This means the detailed dynamical behavior of a classical system
        is very important to understand quantum chaotic transition.
        We claim that this is the generic behavior of quantum
        transition to chaos with 
        soft chaos senario.
        We think this sensitive dependence of quantum spectral measures on
        classical dynamics is a genuine manifestation of quantum chaos.
        This also fortifies that spectral analysis is the legitimate method to
        study quantum chaos.

        The authors would like to thank the other GTP members, 
        C.S. Park, D.H. Yoon,
        S.K. Yoo, and D.K. Park, for informative discussions.

\begin{figure}
\caption{ A typical shape of the one-parameter billiards. $R$ is the
radius of the arc boundaries.   }  
\end{figure}
\begin{figure}
\caption{ The Birkhoff surface of section for $\lambda = 10^{-10}$ after
100,000 iterations. The initial point is given at the neighborhood of 
the boundary of the period two island.  }  
\end{figure}
\begin{figure}
\caption{ Plot for the Brody's exponent $\nu$ versus the parameter
 $\lambda$  }  
\end{figure}
\newpage
\epsfysize=20cm \epsfbox{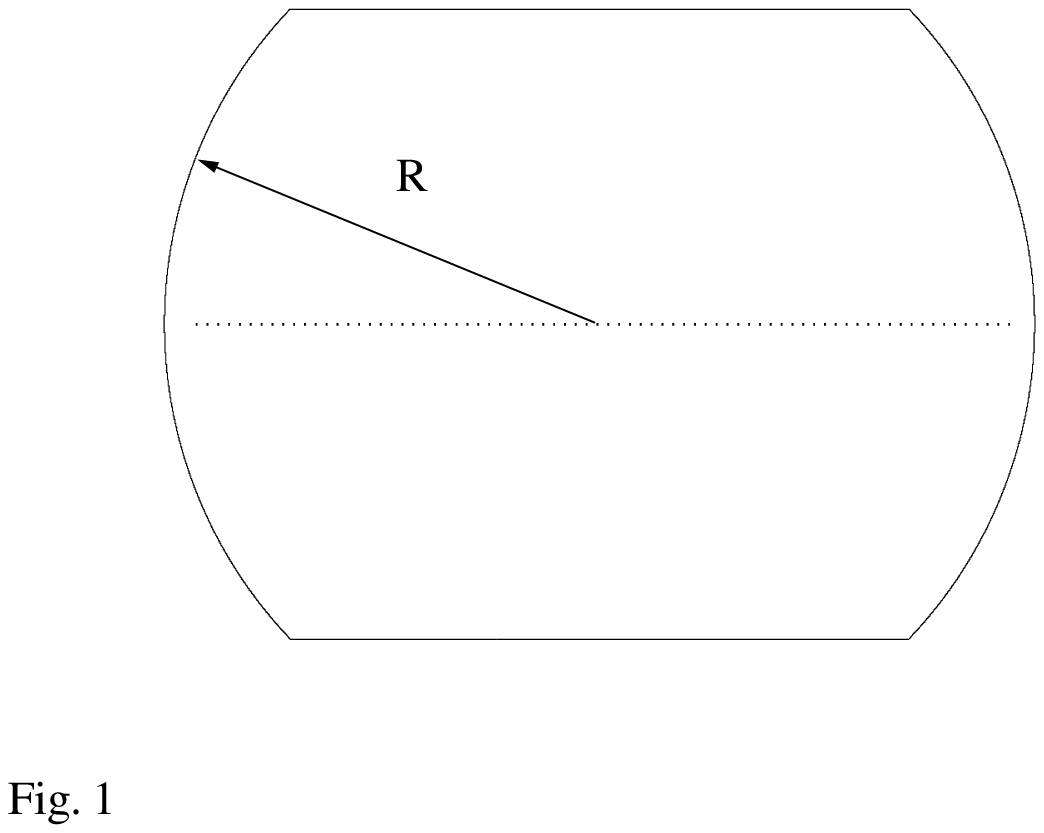}
\newpage
\epsfysize=20cm \epsfbox{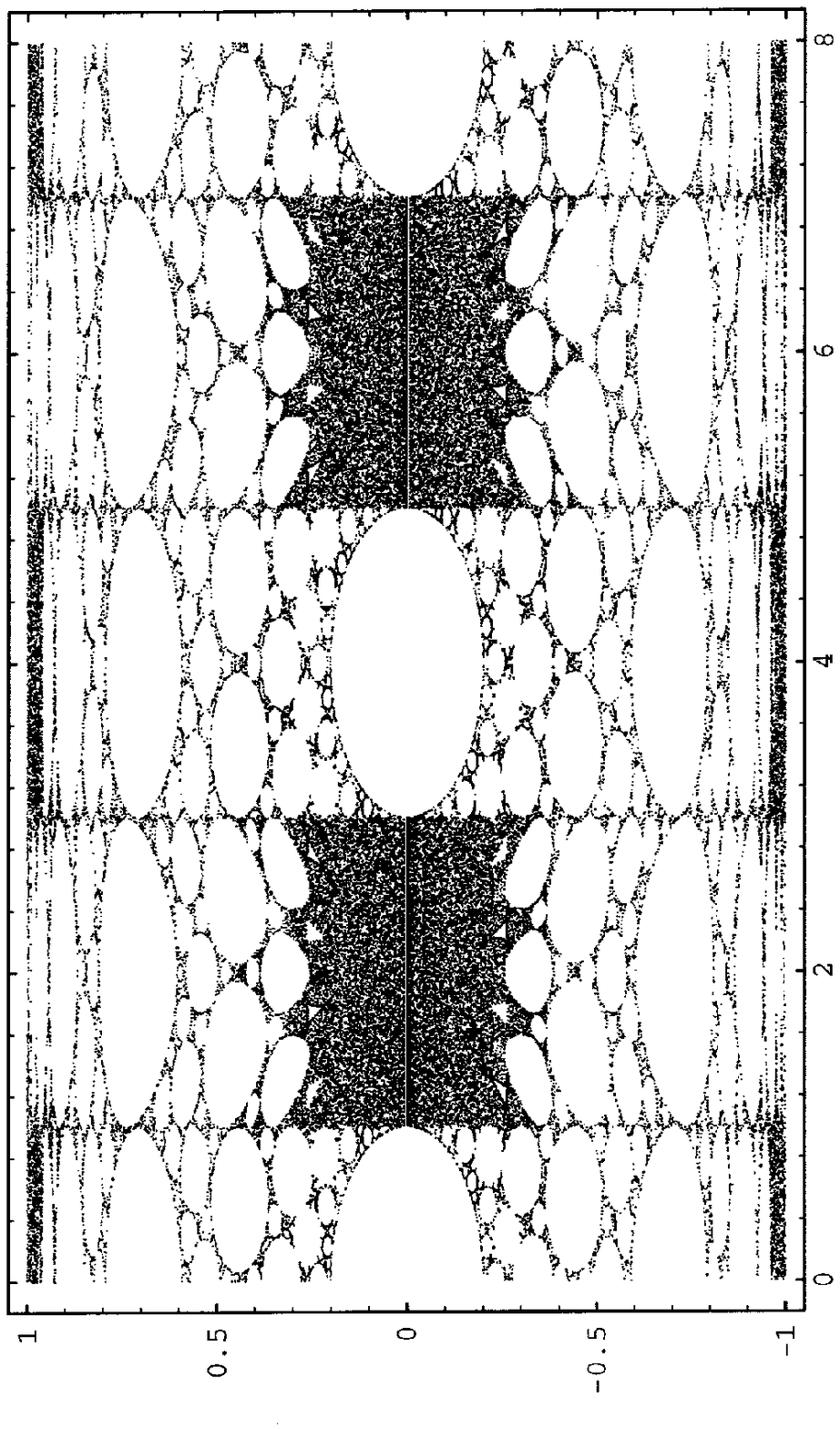}
\newpage
\epsfysize=20cm \epsfbox{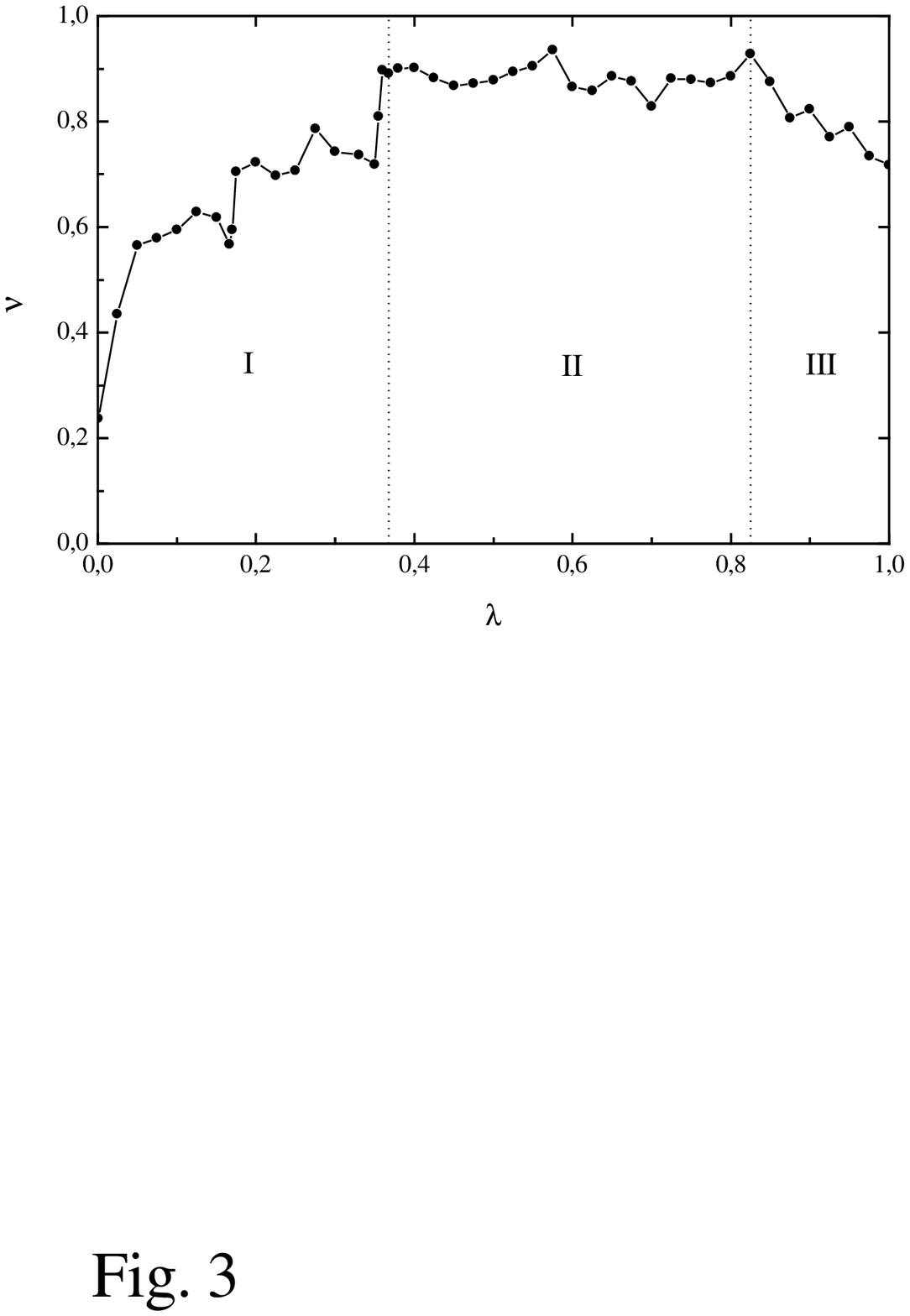}


\begin{thebibliography}{99}
\bibitem{Dt89} As far as one of the authors knows, 
               these billiards are suggested
               to study by Prof. J.F. Dreitlein in the late 
               1989 for the first time.
\end{thebibliography}
\end{document}